\documentclass[final,5p,times,twocolumn,authoryear]{elsarticle}
\usepackage{amssymb}
\usepackage{lipsum}
\usepackage{url}

\usepackage{caption}
\usepackage{subcaption}

\begin{document}

\begin{frontmatter}

\title{BROWDIE: a New Machine Learning Model for Searching T\&Y Dwarfs \\Using the UKIDSS J, H, K Band Survey}

\author[1]{Gwujun Kang\corref{cor}}
\ead{243301@dg1s.hs.kr}
\author[1]{Jiwon Lim}
\author[1]{Bohyun Seo}

\cortext[cor]{Corresponding author}

\affiliation[1]{organization={Daegu Il Science Highschool},
            addressline={987, Gyeongan-ro}, 
            city={Dong-gu},
            postcode={41063}, 
            state={Daegu},
            country={Republic of Korea}}

\begin{abstract}
We propose a new T, Y dwarf search model using machine learning (ML), called the "BROWn Dwarf Image Explorer (BROWDIE). 
Brown dwarfs (BD) are estimated to make up 25 percent of all celestial objects in the Galaxy, yet only a small number have been thoroughly studied. Homogeneous and complete samples of BDs are essential to advance the studies. However, due to their faintness, conducting spectral studies of BDs can be challenging. T\&Y brown dwarfs, a redder and fainter subclass of BDs, are even harder to detect. As a result, only a few T\&Y dwarfs have been extensively studied. Numerous attempts, including ML using various color band observations, have been made to identify BDs based on their colors. However, those models often require a large number of color observations, which can be a limitation. This study implemented an ML model by utilizing data from the J, H, and K photometry of the UKIRT Infrared Deep Sky Surveys (UKIDSS) simultaneously, effectively distinguishing celestial objects similar to BDs. The BROWDIE model was trained using UKIDSS data, and based on the model, 118 T dwarfs and 14 Y dwarfs were found in the UKIDSS DR11PLUS LAS L4 zone.

\end{abstract}

\end{frontmatter}

\section{Introduction}
\label{introduction}

Brown dwarfs (BD) are darkly luminous bodies that do not cause nuclear fusion. They were first predicted in 1963 \citep{kumar1963structure} and observed in 1995 \citep{rebolo1995discovery, nakajima1995discovery}, and they are still being explored and systematically studied. In the spectral classification, BDs occupy spectral types L, T, and Y \citep{avdeeva2023machine}.

There are 25 to 100 billion BDs in the Galaxy \citep{muvzic2017low}, which is about 25\% of the celestial objects in the Galaxy. Despite the large number, they are exceptionally darkly luminous bodies, making their detection and classification complicated.

Regardless of their complex classification, BDs occupy the boundary between stars and planets, and studying their properties helps to refine our understanding of this boundary. A complete and uniform catalog would enable a better determination of the lower mass limit for star formation and the upper mass limit for planet formation. Moreover, BDs share similarities with giant exoplanets, making them valuable analogs for studying exoplanetary atmospheres \citep{avdeeva2023machine}.

There are two main methods of finding BDs: spectroscopy and photometry. Spectroscopy is essential for identifying the properties of BDs and studying their detailed properties. However, spectroscopic observations of a large number of celestial objects across the sky require significant time and resources. Photometry, on the other hand, can cover a much larger area of the sky and capture data from numerous celestial objects simultaneously, making it more efficient \citep{avdeeva2023machine}.

In general, spectroscopy is widely used in BDs searches  because it provides detailed information about their properties. However, spectroscopy has limitations for larger areas. Therefore, photometry is often used in conjunction with spectroscopy. This method searches for BD candidates in a specific area by utilizing the search for a large area, which is the advantage of photometry, and observes spectroscopy only for these candidates, thereby saving time and resources.

For instance, \cite{skrzypek2016photometric} classified 1361 L \& T dwarfs with $J\leq17.5$ from an effective area of $3070\deg^2$ using izYJHKW1W2 photometry from Sloan Digital Sky Survey (SDSS), UKIRT InfraRed Deep Sky Surveys (UKIDSS) and Widefield Infrared Survey Explorer (WISE), to an accuracy of one spectral sub-type. Likewise, \cite{carnero2019brown} presented a catalog of 11,745 photometrically classified BDs with spectral types ranging from L0 to T9 using data from the Vista Hemisphere Survey (VHS) DR3 and the Dark Energy Survey (DES) year 3 release consistent with WISE data, covering $\approx 2400 deg^2$ up to $i_{AB}=22$.

Machine learning (ML) methods enable the search for BDs by identifying correlations that are unique to them in a multidimensional color space. This approach can be particularly useful when color distributions overlap among different celestial objects. As the amount of observational data continues to grow, ML methods are increasingly employed to classify celestial objects.

This study aims to overcome these limitations through ML. The model to be developed, BROWDIE: BROWn Dwarf Image Explorer, is expected to increase the accuracy and efficiency of BD searches by using only three photometric data, JHK from UKIDSS. Unlike the existing BD search model, BROWDIE is expected to expand the search area because it requires only three photometric measurements. This model aims to develop a new search model to distinguish celestial bodies such as quasars with color indices similar to BDs, even when limited photometry data are provided.

\section{Dataset and Preprocessing}
\label{Dataset and Preprocessing}

A dataset was built to train BROWDIEusing data from Simbad, an astronomical database \citep{mukherjee2023picaso}, and PICASO 3.0, a BD climate model \citep{wenger2000simbad}. Data from Simbad was used to create negative class objects. Although many celestial objects presented in Simbad are well studied and with proven spectral type, there are few Y dwarfs available on Simbad — this results in the application of PICASO 3.0. PICASO 3.0 is a one-dimensional climate model for giant planets and BDs. Its application allows for delicate adjustments to BDs and enables the acquisition of detailed spectra in the desired wavelength bands rather than relying solely on photometric ratings.

BROWDIE is specially designed to find T\&Y dwarfs. In general, L/T Transition, the temperature at which L dwarfs and T dwarfs are distinguished, is generally considered to be around 1300K \citep{burrows2006and}. For convenience, this study set the boundary between L dwarfs and T\&Y dwarfs at an effective temperature $(T_{eff} ) = 1300 K$, and  implemented the search model to find BDs with temperatures lower than this boundary. The UKIRT Deep Sky Survey \citep{lawrence2007ukirt} was used for field data, and the transmission curve of the photometer is shown in Figure 1 of \citep{hewett2006ukirt}.

\begin{figure*}
    \centering 
    \includegraphics[width=0.6\textwidth, angle=0]{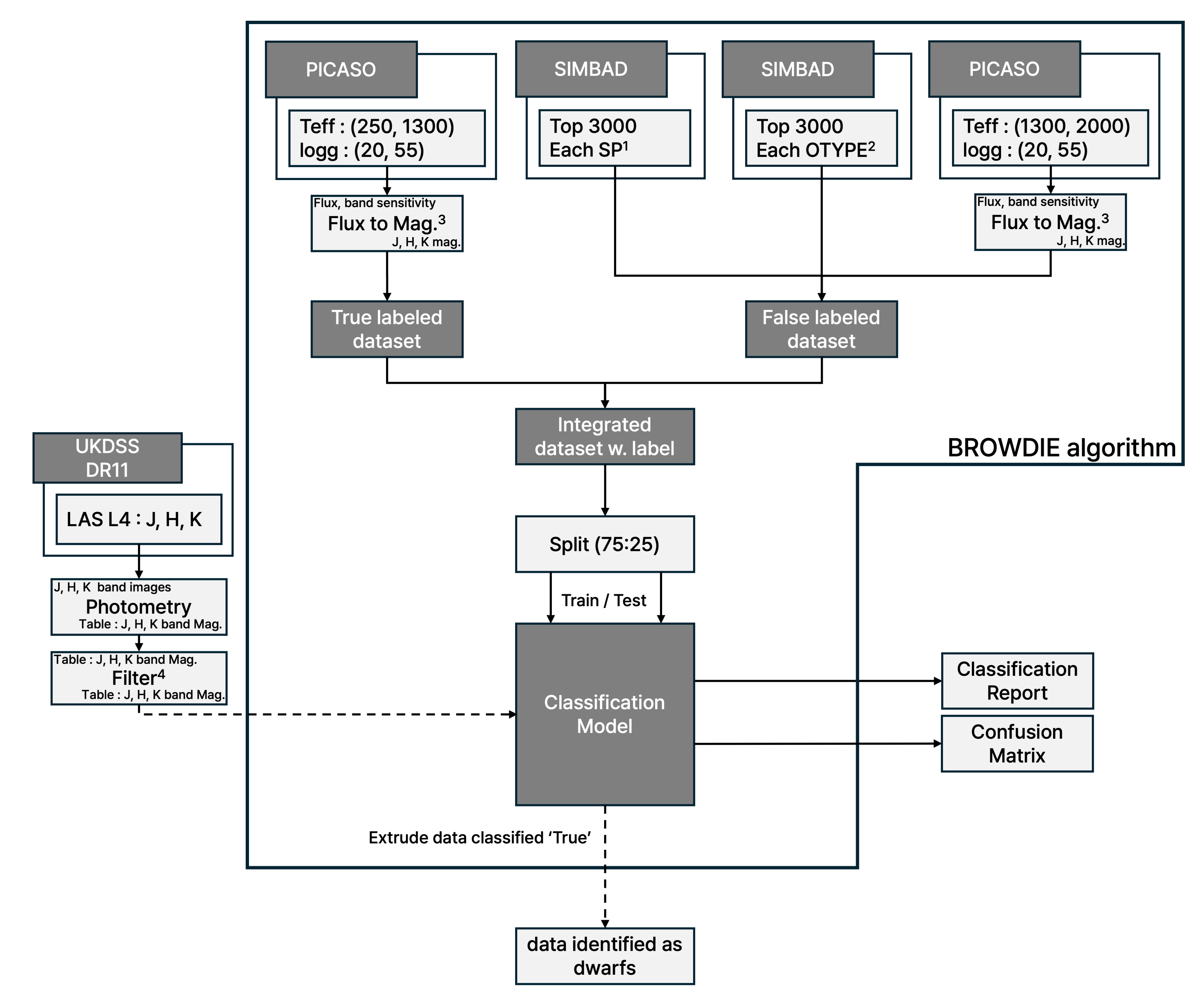}	
    \caption{Overview of the data integration process for the BROWDIE training dataset. The figure illustrates how positive labeled and negative labeled datasets are generated and merged. The positive labeled data, representing T\&Y dwarfs, is obtained through simulated spectra using the PICASO 3.0 model. In contrast, the negative labeled data, representing non-T\&Y dwarfs, is derived from SIMBAD and supplemented with simulated L dwarfs from PICASO 3.0. Both datasets are combined into a single data frame for further processing in the training pipeline. 1 : O, B, A, F, G, K, and M spectral types. 2 : Object type. Data were ordered by \texttt{ident.id} of SIMBAD, for top 3000 items that has all J, H, K band measurements. 3 : Integral of $Flux * Band$ sensitivity, using flux-AB magnitude ($m_{AB}$) conversion to get Mag. Conversion procedure is shown in the \ref{appendix_1}. 4 : Filter : all stars for which Gaia DR3 matches are available.}
\label{fig:BROWDIE_Structure}
\end{figure*}

The method of obtaining the training data was varied differently depending on the positive/negative label assigned to each row of the training data. The whole process is visualized in Figure \ref{fig:BROWDIE_Structure}.

\subsection{Positive Labeled Dataset}
\label{positive labeled dataset}

An object with a positive label means a T\&Y dwarf. A spectrum showing the flux for each wavelength of a simulated T\&Y dwarf was first obtained through PICASO 3.0. The simulated T\&Y dwarf was set as $T_{eff} = [250K, 1300 K]$ and $log\_g = 2.5, 5.5]$, with intervals of $\delta T_{eff} = 10 K$ and $\delta log\_g = 0.1$. This spectrum was then multiplied by the transmission curve of each photometric system of the WFCAM, and integrated to obtain the spectral radiance of a simulated star photographed with a specific photometric system. This procedure is visualized in \ref{appendix_1}. The spectral radiance of the star obtained in this way was converted to a $m_{AB}$ grade and stored.

\subsection{Negative Labeled Dataset}
\label{negative labeled dataset}

An object with a negative label refers to an object that is not a T\&Y dwarf. Up to 3000 of each existing object type, or \texttt{\%OTYPE} in SIMBAD, among celestial objects whose grades J, H, and K can be searched in SIMBAD. \texttt{Ident.id} was used to sort and select the top 3000 object which met the criteria. The Same process was applied to O, B, A, F, G, K, and M spectral types, or \texttt{\%SP} in SIMBAD, under the same conditions.

In addition, PICASO 3.0 was used to effectively identify non-T\&Y dwarfs by adding simulated L dwarfs, convert them to $m_{AB}$ grades in the same way as described above, and storing them.

Both the positive and negative labeled data were merged into a single data-frame for processing in the next section.

\section{Implementation of Machine Learning Models}
\label{Implementation of Machine Learning Models}

This section delves deeper into the application of machine learning in our study. We utilized packages such as TensorFlow \citep{tensorflow2015-whitepaper} and Scikit-learn \citep{scikit-learn} to apply ML techniques to identify BDs. By inputting color index trends into the corresponding package, we obtained information about the presence or absence of BDs through machine learning analysis. In this context, the presence or absence of BDs is our target variable, predicted based on the magnitudes recorded in each photometric system. These magnitudes are directly linked to the color indices, forming the basis of our predictive model.

We applied classifier models using k-NN and RF by Scikit-learn, and Multi-Layer Perceptron (MLP) by Tensorflow. These models were evaluated and compared after training.

To improve ML, magnitudes from each photometric system were converted into color indices and put into the model as part of the preprocessing. Thus, in the JHK photometric system, we obtained J-H, H-K, and J-K color indices.

\subsection{k-Nearest Neighbor}
\label{k-Nearest Neighbor}

A model was built and tested using k-NN. With a 75:25 split ratio for the train and test sets, the model was evaluated. To reduce errors caused by distribution differences during the split process, the model learning and evaluation process was repeated 100 times. The resultant precision, recall, and f1 scores were displayed as a box-and-whisker plot as shown in the first three columns in Figure \ref{fig:intgr bw plot}.

The k-NN model decides prediction positive when the prediction probability is above 20\%. It adopted the number of neighbors of 3.

\subsection{Random Forest}
\label{Random Forest}

Another model was built and tested using RF. Using a split ratio of 75:25 for the train and test sets, respectively, we were able to evaluate the model. The learning and evaluation were repeated 100 times, as in the k-NN model. The resultant precision, recall, and f1 scores were expressed as a box-and-whisker plot, as shown in the second three columns in Figure \ref{fig:intgr bw plot}.

RF model decides prediction positive when the prediction probability is above 20\%.

\subsection{Multi Layer Perceptron}
\label{Multi Layer Perceptron}

MLP was also used to build and test a model. The validation set was re-extracted at a split ratio of $train:validation = 70:30$ from the train set divided by a split ratio of 75:25 between the train set and the test set. Overall, the MLP model used split ratio of $train : validation : test = 52.5 : 22.5 : 25$. The learning and evaluation processes were repeated 20 times for the MLP model.

As in RF, the MLP model decides prediction positive when the prediction probability is above 20\%. In addition, the MLP model adopted 300 epochs and layer number of 3, while dense used 16, 8, and 1, respectively. The Nadam was used for optimizer.

\begin{figure}
    \centering 
    \includegraphics[width=0.4\textwidth, angle=0]{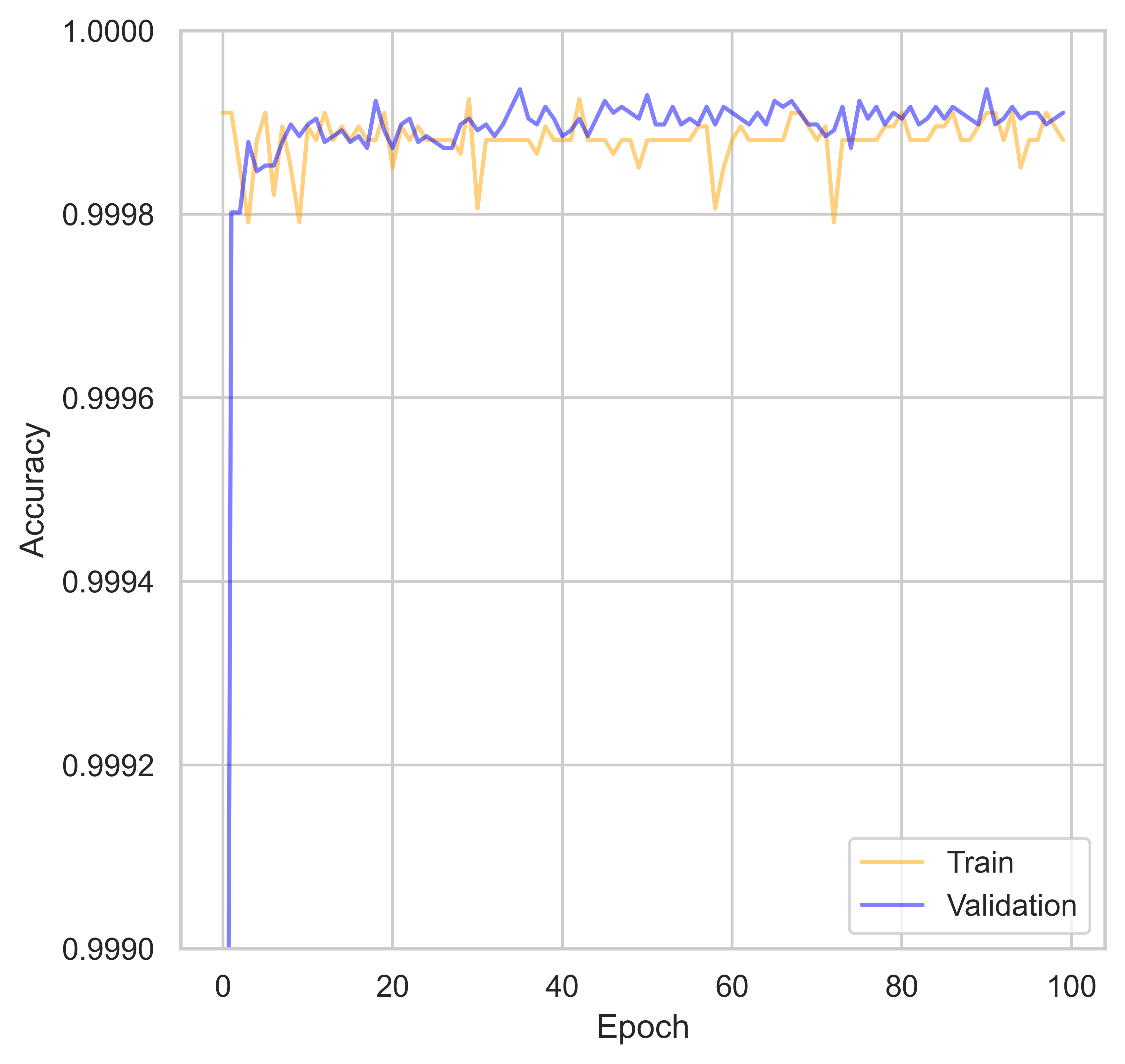}	
    \caption{Randomly selected MLP model's epoch-accuracy graph among 20 } 
\label{fig:epoch-accuracy}
\end{figure}

Figure \ref{fig:epoch-accuracy} presents the epoch-accuracy graph, which is used to assess potential overfitting. The data plotted in the graph represent the performance of a randomly selected MLP model from the evaluation process.

\subsection{Model evaluation}
\label{Model Evaluation}

\begin{figure*}
    \centering 
    \includegraphics[width=0.8\textwidth, angle=0]{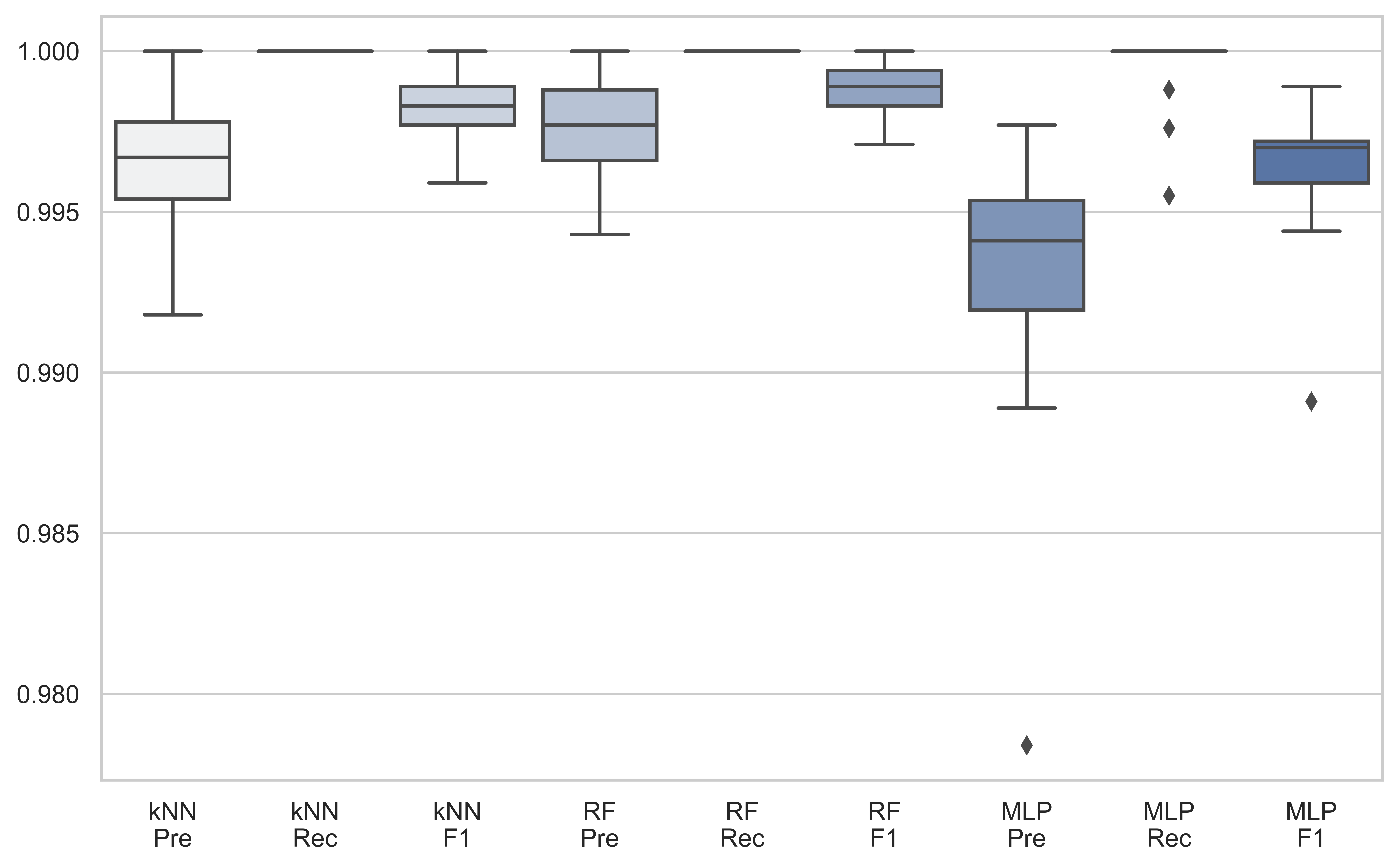}	
    \caption{Box-and-whisker plot comparing the performance of the RF, k-NN, and MLP models used in this study. Precision, recall, and f1 scores for each model were evaluated over 100 repetitions for the RF and k-NN models, and 20 repetitions for the MLP model.} 
\label{fig:intgr bw plot}
\end{figure*}

The resultant precision, recall, and f1 scores are presented as a box-and-whisker plot as shown at the last three column in Figure \ref{fig:intgr bw plot}. The medians of Evaluation indicators are shown in Table \ref{tab:model_comparison}.

\begin{table}
    \centering
    \begin{tabular}{l|ccc}
        \hline
        \textbf{Model} & \textbf{Precision} & \textbf{Recall} & \textbf{F1 score} \\
        \hline
        kNN  & 0.9967 & 1.0    & 0.9983 \\
        RF   & 0.9977 & 1.0    & 0.9989 \\
        MLP  & 0.9941 & 1.0    & 0.997  \\
        \hline
    \end{tabular}
    \caption{Performance comparison of three models. Precision, recall, and f1 scores of each model were evaluated over 100 repetitions for the RF and k-NN models, and 20 repetitions for the MLP model.}
\label{tab:model_comparison}
\end{table}

The Random Forest (RF) model outperformed k-Nearest Neighbor (k-NN) and Multi-Layer Perceptron (MLP) models. As shown in Figure \ref{fig:intgr bw plot}, RF model consistently shows high Precision, Recall, and F1 scores, with a narrow interquartile range (IQR) and median values close to the maximum possible score (i.e., 1.0). Additionally, the RF model exhibits lower variability in performance metrics compared to the MLP model, as indicated by the shorter whiskers in the box plot. The RF model does not show any significant outliers in its performance metrics, whereas the MLP model presents notable outliers, indicating potential instability or inconsistency in its predictions. Overall, the RF model demonstrates reliable performance across various evaluation metrics and is less prone to fluctuations in performance across iterations, making it more suitable in diverse scenarios. 

On the other hand, while k-NN shows high performance, particularly in Recall and F1 scores, the RF model outperforms k-NN in Precision. Therefore, considering all evaluation metrics, the RF model was judged the best overall, and was thus selected as the classification model for BROWDIE.

\section{Application of the BROWDIE Model}
\label{Application of the BROWDIE Model}

\subsection{Production of the Application Data}
\label{Production of the Application Data}

The survey and region selected to find T\&Y dwarfs using the BROWDIE was the UKIDSS DR11PLUS LAS L4. UKIDSS was chosen for its global public release and sufficient precision for BD exploration while covering a wide area. The L4 region of LAS was selected because, at the time of the study, it was close to the meridian at midnight, making observation convenient. Additionally, the region’s rectangular shape was advantageous for the study.

To search for BDs, aperture photometry was first performed on UKIDSS. The resulting photometric magnitudes were then matched to the closest objects within 2 arcseconds in the Gaia Data Release 3 (Gaia DR3) catalog \citep{vallenari2023gaia}. The Gaia DR3 catalog was used as a key to group each object, and the grouped data were then input into the BROWDIE model to identify celestial bodies classified as T\&Y dwarfs.

Using images downloaded from WFCAM, the database that manages UKIDSS images (\url{http://wsa.roe.ac.uk/index.html}), celestial objects were identified, and aperture photometry was performed. The search for celestial objects was conducted using the texttt{daostarfinder} from the Python module texttt{photutils}, and aperture photometry of the detected objects was performed using the texttt{aperture photometry} function of texttt{photutils} \citep{bradley2020astropy}. The inner and outer radii for aperture photometry were set to 14 and 20 pixels, respectively.

\begin{figure}
    \centering
    \includegraphics[width=0.4\textwidth]{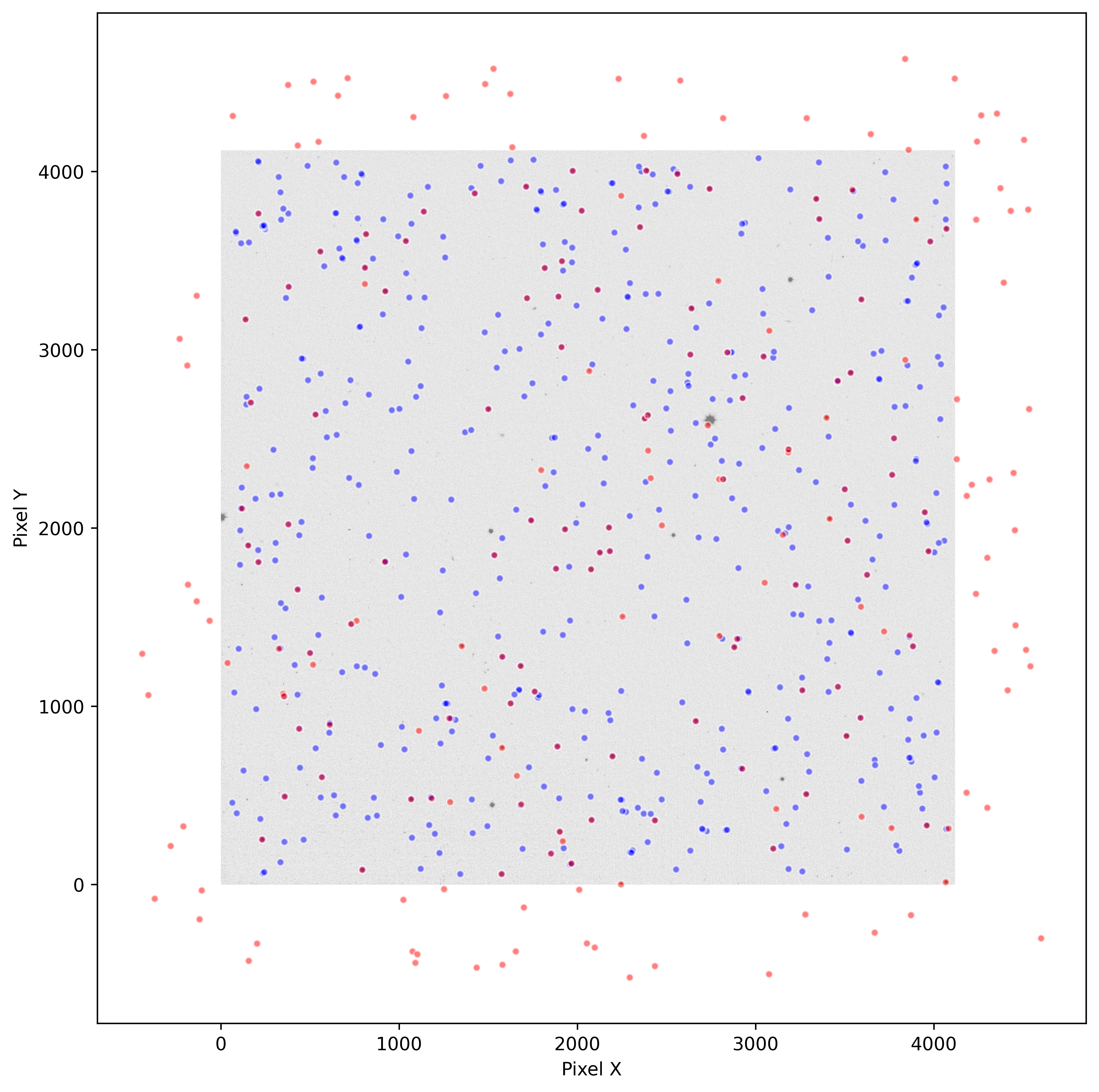}
    \caption{Image of \texttt{w20120502\_00730\_sf\_st.fit HDU 4}. The x-axis of the pixel grid increases in the direction of decreasing Declination (Dec), while the y-axis increases in the direction of decreasing Right Ascension (RA). The image is overlaid with red and blue dots, where red dots represent objects retrieved from the Gaia DR3 catalog within a 1.75 arcminute radius around the image, and blue dots represent objects identified and recognized based on the image. The overlapping of blue and red dots appears as purple.}
\label{fig:w2012_fit}
\end{figure}

The values obtained from aperture photometry were compared to the UKIDSS zeropoint to calculate the magnitude. The right ascension and declination of the photometrically measured objects were determined using the Python module texttt{astropy.wcs}. Subsequently, the celestial data from the Gaia DR3 catalog for objects in the image region were matched with the photometrically measured objects, as shown in Figure \ref{fig:w2012_fit}. At this stage, data were saved only when the celestial distance between the coordinates of the measured object and the Gaia DR3 catalog object was less than 2 arcseconds.

The reason for storing the photometric magnitudes based on the Gaia DR3 catalog is that the Gaia DR3 catalog only records celestial objects beyond the Solar System, preventing the BROWDIE from confusing Solar System minor bodies with T\&Y dwarfs. The objects from the Gaia DR3 catalog and their photometric values were grouped and stored by the object name in the Gaia DR3 catalog. Objects from Gaia DR3 that lacked J, H, or K photometric data were excluded from the BROWDIE application.

Figure \ref{fig:w2012_fit} illustrates the photometric process for the file \texttt{w20120502\_00730\_sf\_st.fit}, part of the UKIDSS DR11PLUS LAS L4. Red dots represent objects retrieved from the Gaia DR3 catalog within a 1.75 arcminute radius around the image, while blue dots represent objects identified and recognized based on the image. The overlap of blue and red dots creates a purple appearance.

\subsection{Application of the BROWDIE Model}
\label{Application of the BROWDIE Model subsection}

The objects selected as the BROWDIE candidates were then analyzed using the BROWDIE to confirm whether they were T\&Y dwarfs. Those identified as T\&Y dwarfs were saved separately.

\subsection{Analysis of the Application}
\label{Analysis of application}

A regression model was developed to estimate the effective temperature $T_{eff}$ using only the data labeled positive in the training set. This model was implemented with Scikit-learn's \texttt{Random Forest Regressor}, and optimal hyperparameters were found using \texttt{GridSearchCV}.

Although some objects with $T_{eff} > 1300K$ were identified in the regression model results, this was attributed to differences between the classifier and regressor models. 

It was observed that objects estimated to have $T_{eff} < 300K$ were significantly distant from the training data points. Based on this observation, objects with $T_{eff} > 1300K$ and $T_{eff} < 300K$ were excluded from the list of T\&Y dwarf candidates. 

To classify the T\&Y dwarf candidates as either T or Y dwarfs, the $T_{eff}$ value of the T/Y transition, which varies among researchers, was set to 470 K, a commonly accepted value. The T\&Y dwarf candidates identified by BROWDIE were then classified into T and Y dwarfs based on the regression model results.

\section{Results and discussion}
\label{Results and Discussion}

In this section, we present and anaylze the outcomes obtained through the BROWDIE model to interpret their significance. This study involved the creation and comparison of three machine learning models: RF, k-NN, and MLP. The RF model, which demonstrated the best performance, was subsequently used to identify T\&Y dwarfs in the UKIDSS DR11PLUS LAS L4 region.

Each model showed excellent performance, but the RF model excelled, consistently achieving high Precision, Recall, and F1 scores, with a narrow interquartile range (IQR) and median values close to the maximum possible scores. The RF model also exhibited lower variability in performance metrics compared to the MLP model which had notable outliers, while the k-NN model showed slightly lower Precision. In summary, the RF model demonstrated reliable performance across various evaluation metrics, with less variation in performance over repetitions, making it the more suitable model for diverse scenarios. Therefore, the RF model was used in further analyses.

Using the BROWDIE model, a total of 132 T and Y spectral type brown dwarfs were identified in the UKIDSS DR11PLUS LAS L4 region, with 118 confirmed as T-type and 14 as Y-type dwarfs. These results indicate that the BROWDIE model can accurately identify celestial objects and generate a meaningful catalog from relevant data, using three photoemtric systems.

In this study, a regression model was developed to predict the $T_{eff}$ of brown dwarfs. This model was implemented using Scikit-learn's Random Forest Regressor, with the best hyperparameters identified through GridSearchCV. The predicted physical properties were used to further filter the T\&Y dwarf candidates. As a result, objects with $T_{eff}$ exceeding 1300K or below 300K were excluded, resulting in a more accurate list of T\&Y dwarf candidates.

However, it is premature to conclude that the model is entirely accurate due to potential omissions, such as quasars. High-redshift quasars, whose visible light is dimmed by the Gunn-Peterson trough, may not be observed by ESA Gaia. Consequently, these high-redshift objects might be excluded during the process of matching photometrically measured objects from UKIDSS with the Gaia DR3 catalog. This initial matching process, intended to exclude solar system bodies, inadvertently led to the omission of quasars as well. Therefore, additional preprocessing techniques are likely needed to improve the model's ability to distinguish quasars.

On the other hand, using data with visible photometric bands or employing a matching process with catalogs like Gaia may help reduce  the misclassification of quasars as brown dwarfs.

Moreover, spectroscopic observations to confirm the false-positive status have not been fully conducted. Nevertheless, the BROWDIE model’s performance metrics suggest that it is effective in searching for T\&Y dwarfs. This study is expected to provide important foundational data for further research related to the spectroscopy of brown dwarfs.

The processed list of T\&Y dwarf candidates can be found in \url{https://github.com/Gwzi/BROWDIE}

\section{Conclusion}
\label{Conclusion}

This study aimed to develop a machine learning-based tool, the BROWDIE (BROWn Dwarf Image Explorer), to distinguish T\&Y dwarfs from photometric data of brown dwarfs. To achieve this, we explored suitable machine learning techniques for dwarf classification and selected three models: k-NN, RF (Random Forest), and MLP (Multi-Layer Perceptron). Each model's performance was evaluated by repeating the training and validation processes 100 times for the k-NN and RF models and 20 times for the MLP model. The results showed that the RF model consistently maintained a high F1 score, leading to its implementation of the BROWDIE based on the RF model.

The trained model identified a total of 132 T\&Y dwarfs in the UKIDSS DR11PLUS LAS L4 region, of which 118 were confirmed as T dwarfs and 14 as Y dwarfs. These results indicate that the BROWDIE model can accurately identify celestial objects and generate a meaningful catalog from related data, using 3 photometric systems.

However, while the BROWDIE model shows promise in effectively searching for T\&Y dwarfs, its accuracy is not yet fully reliable due to potential omissions, such as high-redshift quasars, which might have been excluded during the matching process with the Gaia DR3 catalog, due to ESA Gaia's detectable wavelength range. This exclusion, originally intended to filter out solar system bodies, inadvertently led to the omission of quasars. 

On the other hand, using visible photometric bands or matching with catalogs like Gaia may reduce the misclassification of quasars as brown dwarfs.

Additional preprocessing techniques and possibly incorporating visible photometric data could help mitigate the risk of misclassifying quasars as brown dwarfs. Although spectroscopic confirmation of false positives has not been fully conducted, the model's performance metrics suggest that BROWDIE is a valuable tool for brown dwarf research, offering a solid foundation for further studies in this field.

\section*{Acknowledgements}
\label{Acknowledgements}

This research has made use of the UKIDSS project. The UKIDSS project is defined in \citep{lawrence2007ukirt}. UKIDSS uses the UKIRT Wide Field Camera \citep{casali2007ukirt}. The photometric system is described in \citep{hewett2006ukirt}, and the calibration is described in \citep{hodgkin2009ukirt}. The pipeline processing and science archive are described in \citep{hambly2008wfcam}. 

This research has made use of the SIMBAD database,
operated at CDS, Strasbourg, France.

We would like to express our sincere gratitude to the anonymous referees for their valuable comments and suggestions, which have significantly improved the quality and clarity of this paper. we greatly appreciate the time and effort they dedicated to this process.

\appendix
\section{flux-$m\_AB$ conversion}
\label{appendix_1}

\begin{figure*}
    \centering
    \includegraphics[width=0.7\textwidth]{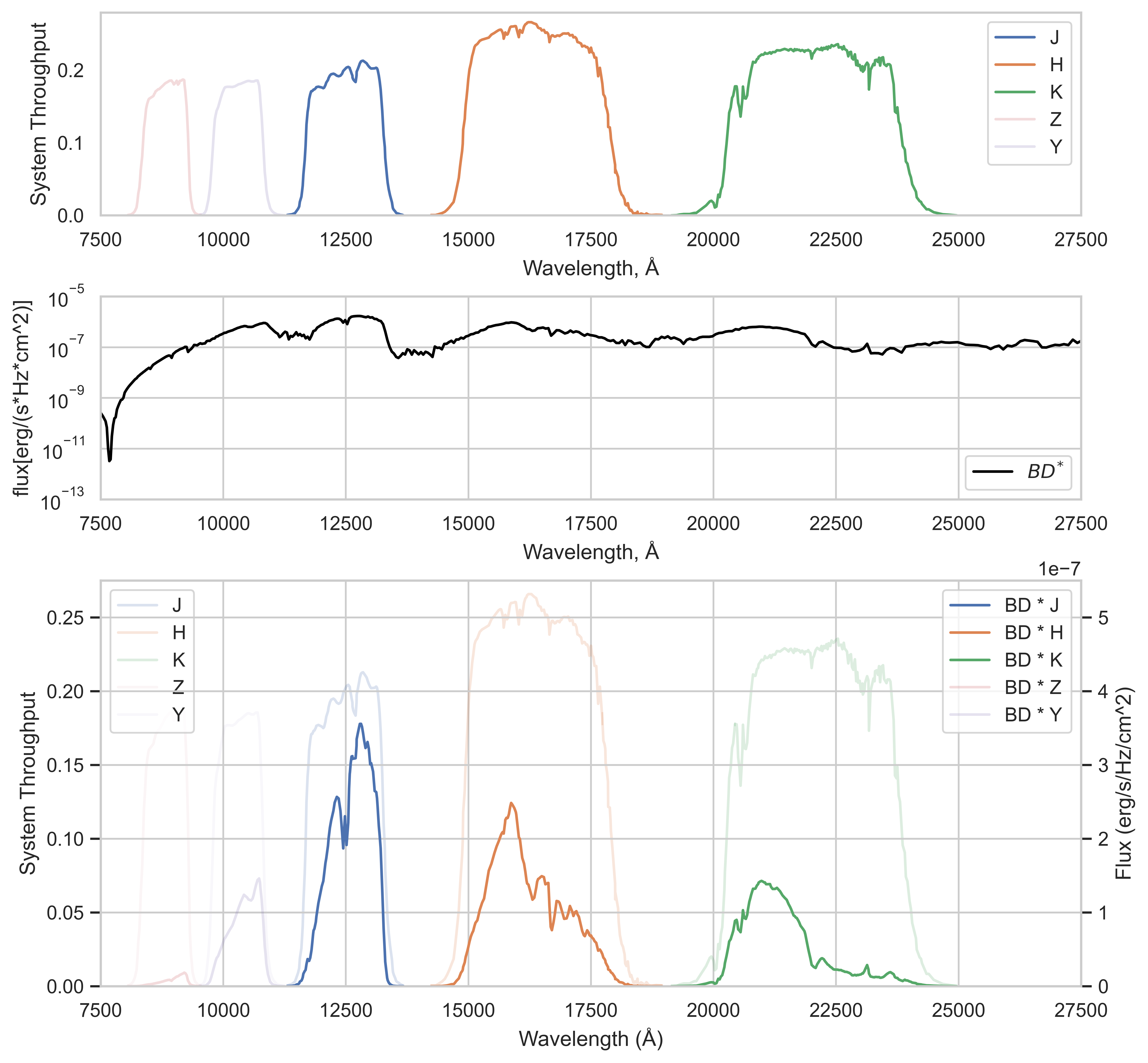}
    \caption{(Top) : System throughput per wavelength of the UKIRT's WFCAM J, H and K bands, which consistitute the photometric system used in this study. The Y and Z bands are shown for comparison. (Center) : Spectrum of BD that has $T_{eff} = 1200K, log\_g = 3.0$, made with PICASO 3.0, for visualization example for flux-$m_{AB}$ conversion procedure. (Bottom) : multiplication of each system throughput of the UKIRT's WFCAM J, H, K bands, with spectrum shown in \textit{Center}. The Y and Z bands are shown for comparison. This spectrum is multiplied by the transmission curve of each photometric system of the WFCAM, and integrated to obtain the spectral radiance of a simulated star photographed with a specific photometric system.}
\label{fig:spectrum calculation}
\end{figure*}

To convert the spectra provided by PICASO 3.0 to $m_{AB}$, it was first necessary to check the spectra that the sensor would acquire after passing through each filter. This calculation was made by multiplying the system throughput of the filter by the desired simulated BD's spectrum. The spectra of the simulated BD object that have passed through the filter thus obtained are shown in the (Bottom) of Figure \ref{fig:spectrum calculation}. Integrating these spectra yields the flux density at which the sensor detects the simulated BD that has passed through the filter. The obtained flux density is compared with the flux density of zero magnitude for each filter star provided by \citep{hewett2006ukirt} to obtain the absolute $m_{AB}$ using Pogson's equation.

\bibliographystyle{elsarticle-harv} 
\bibliography{BROWDIE}

\end{document}